\documentclass[showpacs,showkeys,eqsecnum,twocolumn]{revtex4}

\usepackage{amsfonts,amsmath,amssymb,bm,hyperref,graphicx,accents}

\begin{document}

\preprint{HUPD 0602}

\title{Neutrino spin oscillations in gravitational fields}

\author{Maxim Dvornikov}
\affiliation{Institute of Terrestrial Magnetism,
Ionosphere \\ and Radiowave Propagation (IZMIRAN) \\
142190, Troitsk, Moscow region, Russia} \email{maxdvo@izmiran.ru}
\affiliation{Graduate School of Science, Hiroshima University,
Higashi-Hiroshima, Japan}

\date{\today}

\begin{abstract}
We study neutrino spin oscillations in gravitational fields. The
quasi-classical approach is used to describe the neutrino spin
evolution. First we examine the case of a weak gravitational
field. We obtain the effective Hamiltonian for the description of
neutrino spin oscillations. We also receive the neutrino
transition probability when a particle propagates in the
gravitational field of a rotating massive object. Then we apply
the general technique to the description of neutrino spin
oscillations in the Schwarzschild metric. The neutrino spin
evolution equation for the case of the neutrino motion in the
vicinity of a black hole is obtained. The effective Hamiltonian
and the transition probability are also derived. We examine the
neutrino oscillations process on different circular orbits and
analyze the frequencies of spin transitions. The validity of the
quasi-classical approach is also considered.
\end{abstract}

\pacs{14.60.Pq, 14.60.St, 04.25.Nx, 04.25.-g}

\keywords{neutrino spin oscillations, gravitational fields,
Schwarzschield metric}

\maketitle

\section{introduction}

Neutrino physics becomes one the most interesting fields of
elementary particle physics, especially in the wake of the recent
experimental achievements in studying of solar and reactor
neutrinos (see, e.g., Refs.~\cite{Aha04,Ali05}). One of the most
intriguing puzzles in neutrino physics in the neutrino
oscillations problem. Neutrinos of one type can be converted into
another type if there is a mixing between different neutrino
eigenstates. Presently three major opportunities for neutrino
oscillations are discussed in literature. The first one is
neutrino flavor oscillations. The example of this oscillations
type are $\nu_e\to\nu_{\mu,\tau}$ or $\nu_\mu\to\nu_\tau$
conversions that are the possible explanations of the solar and
atmospheric neutrino problems. The second type of neutrino
oscillations is the transitions between neutrino helicity states
within one flavor, or neutrino spin oscillations. The last
opportunity is the combination of the two first neutrino
oscillations types, i.e. the transition when both flavor and
helicity states can change. This type of neutrino oscillations is
called neutrino spin-flavor oscillations. It can be important in
solving of the solar neutrino problem. In this paper we restrict
ourselves to studying of neutrino spin oscillations.

It was realized more than twenty years ago that external fields
drastically change the process of vacuum neutrino oscillations. In
Refs.~\cite{Wol78,MikSmi85eng} it was established that neutrino
weak interactions with background matter result in the resonant
enhancement of neutrino flavor oscillations. The electromagnetic
interactions are of great importance in studying of neutrino spin
and spin-flavor oscillations. For example, the neutrino
interaction with an external electromagnetic field provides one of
the mechanisms for the mixing between different helicity
eigenstates. It is also interesting to study the influence of
gravitational fields on neutrino oscillations (neutrino does not
seem to participate in strong interactions). Despite the
gravitational interaction is relatively weak compared to
electromagnetic and weak interactions there are rather strong
gravitational fields in the Universe. Thus neutrino oscillations
in gravitational fields are of interest in astroparticle physics
and cosmology.

The influence of the gravitational interaction on neutrino
oscillations has been studied in many publications (see, e.g.,
Refs.~\cite{CasMon94,AhlBur96,PirRoyWud96,Ahl97,CarFul97,KosMew04,SinMobPap04,DvoGriStu05}).
First we mention Refs.~\cite{PirRoyWud96,CarFul97} where the
formalism for the description of neutrino oscillations in
gravitational fields in presence of matter and external
electromagnetic fields was worked out. One of the approaches to
the description of the spin dynamics of a neutrino in a
gravitational field consists in decomposing of the Dirac
Hamiltonian (in presence of an external gravitational field) and
establishing terms that mix different \emph{chirality} components
of the neutrino wave function. This method was used in
Ref.~\cite{CasMon94}. To describe the \emph{helicity} evolution in
a gravitational field one can use the evolution equation in
Heisenberg representation with the Hamiltonian in Foldy-Wouthuysen
representation. This approach was implemented in
Ref.~\cite{SinMobPap04}. Neutrino spin oscillations and spin light
of a neutrino in weak gravitational fields were studied in our
work~\cite{DvoGriStu05}. Gravity induced neutrino flavor
oscillations were examined in Refs.~\cite{AhlBur96,Ahl97}.

There is, however, another approach for the treatment of the
spinning particle dynamics in an external gravitational field. The
system of equations for the description of the spinning particle
motion in gravitational fields was derived by A.~Papapetrou in
Ref.~\cite{Pap51}. Since then plenty of works where Papapetrou
equations were analysed have been published. The major
difficulties in solving Papapetrou equations are that these
equations are non-linear in the particle spin as well as the
motion of a spinning particle deviates from the geodesical. Thus
to describe the motion of a \emph{finite size} spinning particle
in a gravitational field one has to take into account the
contribution of the spin term into the particle motion law. It was
established in Ref.~\cite{KhrPom98} that Papapetrou equations can
be significantly simplified when one uses the quasi-classical
approach and linear in spin approximation of these equations for a
\emph{point} particle.

In this paper we study neutrino spin oscillations in gravitational
fields within the quasi-classical approach. Note that the
quasi-classical treatment of neutrino spin oscillations in moving
and polarized matter under the influence of electromagnetic fields
was considered in Refs.~\cite{EgoLobStu00,LobStu01,DvoStu02JHEP}.
In Sec.~\ref{General} we formulate the main equations necessary
for the description of the neutrino motion and spin evolution. The
limit of a weak gravitational field is examined in Sec.~\ref{WGF}.
On the basis of the general equation presented in
Sec.~\ref{General} we obtain the effective Hamiltonian for
neutrino spin oscillations. This effective Hamiltonian is valid
for arbitrary neutrino velocity. We also compare the neutrino spin
evolution equation with the analogous one derived in our previous
work \cite{DvoGriStu05} and find out that the equation obtained in
the present work correctly accounts for the neutrino velocity
dependence. Then we receive neutrino transition probability when a
particle propagates in the gravitational field of a rotating
massive object. In Sec.~\ref{SM} we apply the general technique to
the description of neutrino spin oscillations in the Schwarzschild
metric. We obtain the neutrino spin evolution equation which is
valid for the neutrino motion even in the vicinity of a black
hole. The effective Hamiltonian and the transition probability are
also derived. We examine neutrino oscillations process on
different circular orbits and analyze the frequencies of the spin
transitions. In Sec.~\ref{CONCL} we discuss our results. The
calculation of covariant derivatives of the vierbein vectors
(Appendix~\ref{VF}) and the basic elements of $SL(2,C)$ group
(Appendix~\ref{LorGroup}) are also presented.

\section{Motion of a spinning point particle in
gravitational fields}\label{General}

The motion of a spinning particle in gravitational fields was
described in Ref.~\cite{Pap51}. The evolution of the spin tensor
of a particle $S^{\mu\nu}$ and particle's momentum $p^\mu$ is
described by the following equations,
\begin{align}
  \label{PapS}
  \frac{DS^{\mu\nu}}{D\tau} & =
  p^{\mu}v^{\nu}-p^{\nu}v^{\mu}, \\
  \label{Papp}
  \frac{Dp^{\mu}}{D\tau} & =-\frac{1}{2}
  R^{\mu}{}_{\nu\rho\sigma}v^\nu S^{\rho\sigma},
\end{align}
where $v^\mu$ is the unit tangent vector to the center-of-mass
world line, $\tau$ is the parameter (not necessarily the proper
time) which changes along the world line, $D/D\tau$ denotes the
covariant derivative along the world line and
$R^{\mu}{}_{\nu\rho\sigma}$ is the Riemann tensor. The spin vector
is introduced in the following way (see, e.g., Ref.~\cite{Wal72}),
\begin{equation*}
  S_\rho=\frac{1}{2m}\sqrt{-g}
  \varepsilon_{\mu\nu\lambda\rho}p^\mu S^{\nu\lambda},
\end{equation*}
where $\varepsilon_{\mu\nu\lambda\rho}$ is the completely
antisymmetric tensor density, $g=\det(g_{\mu\nu})$ and $m^2=p_\mu
p^{\mu}$ is the constant of the particle's motion.

When we study the motion of point particles in a gravitational
field, Eqs.~\eqref{PapS} and \eqref{Papp} can be significantly
simplified. It was shown in Ref.~\cite{Wei72p121} that using only
the Principle of General Covariance the equations for the
description of particle's spin vector $S^{\mu}$ and four-velocity
$U^\mu=dx^\mu/d\tau$ evolution take the form,
\begin{align}
  \label{WeiS}
  \frac{DS^\mu}{D\tau} & =0, \\
  \label{WeiU}
  \frac{DU^\mu}{D\tau} & =0.
\end{align}
Here the covariant derivatives are taken with respect to the
proper time $\tau$. It follows from Eqs.~\eqref{WeiS} and
\eqref{WeiU} that particle's spin and four-velocity are parallel
transported along its world-line. It should be noted that usual
properties of the spin vector $S^\mu$ in Minkowski space-time,
namely $S^\mu U_\mu=0$ and $S^\mu S_\mu$ is the constant value,
remain unchanged when a particle moves in curved space-time. One
can verify these properties using Eqs.~\eqref{WeiS} and
\eqref{WeiU}. The basic Eqs.~\eqref{WeiS} and \eqref{WeiU} can be
also rewritten with help of the Christoffel symbols of the second
kind $\Gamma^{\mu}{}_{\alpha\beta}$,
\begin{align}
  \label{SGamma}
  \frac{dS^\mu}{d\tau} & =
  -\Gamma^{\mu}{}_{\alpha\beta}U^{\alpha}S^{\beta}, \\
  \label{UGamma}
  \frac{dU^\mu}{d\tau} & =
  -\Gamma^{\mu}{}_{\alpha\beta}U^{\alpha}U^{\beta}.
\end{align}

Eq.~\eqref{SGamma} describes the spin evolution in a general
coordinate system. However the particle's properties are
determined by the spin vector components measured with respect to
the particle's rest frame. The particle's rest-frame-spin
precession in gravitational fields was discussed in
Ref.~\cite{KhrPom98}. In order to proceed in our analysis of
particle's spin evolution, we should rewrite Eqs.~\eqref{SGamma}
and \eqref{UGamma} in a locally minkowskian frame. One can
implement this coordinates transformation with help of the
vierbein vectors $e^a{}_\mu$. We recall the basic vierbein vectors
properties,
\begin{align}\label{vierbeinprop}
  g_{\mu\nu} & =e^a{}_\mu e^b{}_\nu\eta_{ab},
  \quad
  \delta^\mu{}_\nu=e_a{}^\mu e^a{}_\nu,
  \\
  \notag
  \eta_{ab} & =e_a{}^\mu e_b{}^\mu g_{\mu\nu},
  \quad
  \delta^a{}_b=e^a{}_\mu e_b{}^\mu,
\end{align}
where $e_a{}^\mu=\eta_{ab}g^{\mu\nu}e^b{}_\nu$ is the inverse
vierbein and $\eta_{ab}=\mathrm{diag}(+1,-1,-1,-1)$ is the metric
tensor in a locally minkowskian frame.

The components of the spin and four-velocity with respect to a
locally minkowskian frame are,
\begin{equation*}
  s^a=e^a{}_\mu S^\mu,
  \quad
  u^a=e^a{}_\mu U^\mu.
\end{equation*}
The equations for the description of the $s^a$ and $u^a$ evolution
have the form (see also Ref.~\cite{KhrPom98}),
\begin{align}
  \label{sG}
  \frac{ds^a}{dt} & =
  \frac{1}{\gamma}G^{ab}s_b, \\
  \label{uG}
  \frac{du^a}{dt} & =
  \frac{1}{\gamma}G^{ab}u_b,
\end{align}
where $G^{ab}=\eta^{ac}\eta^{bd}\gamma_{cde}u^e=-G^{ba}$ is the
analog of the electromagnetic field tensor and
$\gamma_{abc}=\eta_{ad}e^d{}_{\mu;\nu}e_b{}^{\mu}e_c{}^{\nu}$ are
the Ricii rotation coefficients. In Eqs.~\eqref{sG} and \eqref{uG}
we keep the usual notation $\gamma=U^0=dt/d\tau$, however it
should be noted that $\gamma\not= u^0$. The derivatives in the
right-handed sides of these equations are taken with respect to
the laboratory time.

To make the coordinate transformation into the particle's rest
frame one uses a boost within a locally minkowskian frame. The
relation between the spin vector $s^a$
and the three dimensional spin in the particle's rest frame
$\bm{\zeta}$ is given by following expression,
\begin{equation}\label{szeta}
  s^a=
  \left(
    (\bm{\zeta}\mathbf{u}),\bm{\zeta}+
    \frac{\mathbf{u}(\bm{\zeta}\mathbf{u})}{1+u^0}
  \right).
\end{equation}
Here $u^0$ and $\mathbf{u}$ are the time and space components of
the four-velocity $u^a$. We remind that $u^a$ is the four velocity
vector in the vierbein frame.

Using Eqs.~\eqref{sG} and \eqref{szeta} we can readily derive the
equation for the description of the three dimensional spin vector
evolution,
\begin{equation}\label{zetaeq}
  \frac{d\bm{\zeta}}{dt}=
  \frac{2}{\gamma}[\bm{\zeta}\times\mathbf{G}],
\end{equation}
where
\begin{equation}\label{GEB}
  \mathbf{G}=\frac{1}{2}
  \left(
    \mathbf{B}+
    \frac{1}{1+u^0}[\mathbf{E}\times\mathbf{u}]
  \right).
\end{equation}
In deriving of Eqs.~\eqref{zetaeq} and \eqref{GEB} we use the fact
that any antisymmetric tensor in four dimensional (minkowskian)
space-time can be expressed in terms of the two three dimensional
vectors (analogs of electric and magnetic fields), i.e.
$G_{ab}=(\mathbf{E},\mathbf{B})$, where $G_{0i}=E_i$ and
$G_{ij}=-\varepsilon_{ijk}B_k$. Eqs.~\eqref{zetaeq} and
\eqref{GEB} describe particle's spin precession in arbitrary
gravitational field. These equations are linear in spin vector.
One can, however, discuss further (non-linear in spin) terms
contributing to the particle's spin evolution (see
Ref.~\cite{KhrPom98}).

At the end of this section me mention that Eqs.~\eqref{sG} and
\eqref{uG} [or alternatively Eqs.~\eqref{zetaeq} and \eqref{GEB}]
are similar to the equations for the description of the spin and
four-velocity evolution of a charged particle with gyromagnetic
ratio $\mathfrak{g}=2$ interacting with an external
electromagnetic field (see also Ref.~\cite{KhrPom98}). The spin
evolution of a charged particle with $\mathfrak{g}=2$ was studied
in Refs.~\cite{Zwa65,LobPav99}. It was shown that the spin
dynamics was completely determined by the particle's motion law,
i.e. by the solution of the Lorentz equation \eqref{uG}.

\section{Neutrino spin evolution in weak gravitational fields}
\label{WGF}

In this section we study the particle spin precession in a weak
gravitational field. We describe the neutrino spin oscillations
and discuss the obtained results as well as we consider analogous
approach elaborated in our previous work.

Weak gravitational fields can be found, for instance, at great
distances from a finite massive object under study. In this case
we can always choose the quasi-minkowskian coordinate system,
\begin{equation}\label{weakgf}
  g_{\mu\nu}=\eta_{\mu\nu}+h_{\mu\nu},
\end{equation}
where $\eta_{\mu\nu}=\mathrm{diag}(+1,-1,-1,-1)$ is the Minkowski
metric tensor. The metric perturbation $h_{\mu\nu}$ has to vanish
at the infinity. One of the possible examples of the metric given
in Eq.~\eqref{weakgf} is the gravitational field created by a
massive rotating object at great distances. Using post-Neutonian
approximation we obtain for the components of the tensor
$h_{\mu\nu}$ (see, e.g., Ref.~\cite{Wei72p212}),
\begin{gather}
  \label{hmunuPN}
  h_{00}=2\varphi,
  \quad
  h_{ij}=2\varphi\delta_{ij},
  \quad
  h_{0i}=-h_i,
  \\
  \label{phih}
  \varphi=-\frac{M}{r},
  \quad
  \mathbf{h}=\frac{2}{r^3}
  [\mathbf{r}\times\mathbf{J}],
\end{gather}
where $M$ is the mass of the object, $\mathbf{J}$ is its total
angular momentum and $r$ is the distance from the object.

First let us consider the particle's spin evolution when an
observer is in the particle's rest frame. For this purpose we can
use directly Eq.~\eqref{SGamma}. In our case the four dimensional
spin vector is reduced to the three dimensional one, i.e.
$S^\mu\to(0,\bm{\zeta})$. Eq.~\eqref{SGamma}, written for the
spatial components of the spin vector, takes the following form,
\begin{equation}\label{zetaeqrfG}
  \frac{d\zeta_i}{dt}=
  -\Gamma^{i}{}_{0j}\zeta_{j}.
\end{equation}
Here we assume that $U^0=1$ and $\mathbf{U}=0$ in particle's rest
frame. Note that we do not distinguish between upper and lower
indexes when we use three dimensional vectors. Christoffel symbols
$\Gamma^{i}{}_{0j}$ can be calculated with help of
Eq.~\eqref{hmunuPN},
\begin{equation}\label{Gi0j}
  \Gamma^{i}{}_{0j}=
  \frac{1}{2}
  \left(
    \frac{\partial h_i}{\partial x_j}-
    \frac{\partial h_j}{\partial x_i}
  \right)=
  -\frac{1}{2}\varepsilon_{ijk}
  [\bm{\nabla}\times\mathbf{h}]_k.
\end{equation}
Using Eqs.~\eqref{zetaeqrfG} and \eqref{Gi0j} we obtain the
equation for the particle's spin evolution in its rest frame,
\begin{equation}\label{zetaeqr}
  \frac{d\bm{\zeta}}{dt}=
  \frac{1}{2}[\bm{\zeta}\times[\bm{\nabla}\times\mathbf{h}]].
\end{equation}
One can see that Eq.~\eqref{zetaeqr} coincides (to within the
sign) with the analogous equation derived in our previous work
(see Ref.~\cite{DvoGriStu05}) if we set there $\gamma=1$ and
$\bm{\beta}=0$.

Now let us discuss the particle's spin precession in a weak
gravitational field, with an observer being placed in the
laboratory frame, i.e. a particle has the non-zero velocity with
respect to him. In this case we should use Eqs.~\eqref{zetaeq} and
\eqref{GEB} to describe particle's spin evolution. The Ricci
rotation coefficients were calculated in Ref.~\cite{KhrPom98} in
the weak gravitational field approximation. They can be expressed
in the following way,
\begin{equation}\label{Ricciwf}
  \gamma_{abc}=\frac{1}{2}
  \left(
    \frac{\partial h_{bc}}{\partial x^a}-
    \frac{\partial h_{ac}}{\partial x^b}
  \right),
\end{equation}
where $h_{ab}$ and $x^a$ are the components of the tensor
$h_{\mu\nu}$ and vector $x^\mu$ in the vierbein frame. Note that
we do not distinguish the vierbein and world indexes in the weak
gravitational field approximation.

Using Eqs.~\eqref{hmunuPN} and \eqref{Ricciwf} as well as the
definition of the tensor $G_{ab}$ we receive the expressions for
the "electric" and "magnetic" fields,
\begin{equation*}
  \mathbf{E}=-\gamma\bm{\nabla}\varphi+
  \frac{\gamma}{2}v_i\bm{\nabla}h_i,
  \quad
  \mathbf{B}=\gamma[\mathbf{v}\times\bm{\nabla}]\varphi+
  \frac{\gamma}{2}[\bm{\nabla}\times\mathbf{h}].
\end{equation*}
One can readily find the vector $\mathbf{G}$, which determines the
particle's spin precession, in the explicit form,
\begin{align}\label{Gwf}
  \mathbf{G}= & \frac{\gamma}{2}
  \bigg(
    \frac{1}{2}[\bm{\nabla}\times\mathbf{h}]+
    \frac{2\gamma+1}{\gamma+1}[\mathbf{v}\times\bm{\nabla}]\varphi
    \\
    \notag
    & -
    \frac{\gamma}{2(\gamma+1)}v_i[\mathbf{v}\times\bm{\nabla}]h_i
  \bigg),
\end{align}
where $\mathbf{v}=\dot{\mathbf{r}}$ is the particle's velocity and
$\gamma=(1-v^2)^{-1/2}$. Note that if we set $\mathbf{v}=0$ in
Eq.~\eqref{Gwf}, we obtain the expression consistent with
Eq.~\eqref{zetaeqr} obtained directly from Eq.~\eqref{SGamma}. It
proves the validity of the used technique.

Eqs.~\eqref{zetaeq} and \eqref{Gwf} govern particle's spin
evolution in a weak gravitational field. These equations are valid
for arbitrary particle velocities. Using Eqs.~\eqref{zetaeq} and
\eqref{Gwf} we can describe neutrino spin oscillations in the weak
gravitational field. For example, it is possible to study the
particle's spin dynamics of a neutrino emitted in the vicinity of
a rotating black hole and then propagating faraway from the
massive object (see Ref.~\cite{DvoGriStu05}). Supposing that a
neutrino has the velocity directed along the $\mathbf{e}_z$ base
vector we obtain from Eq.~\eqref{Gwf} the expression for the
effective Hamiltonian (see Ref.~\cite{EgoLobStu00})
\begin{equation}\label{Heffwf}
  H_\mathrm{eff}=-\frac{1}{\gamma}(\bm{\sigma}\mathbf{G}),
\end{equation}
where $\bm{\sigma}=(\sigma_{1},\sigma_{2},\sigma_{3})$ are the
Pauli matrices.

Let us consider neutrino spin oscillations when a neutrino
interacts with the gravitational field having the properties given
in Eq.~\eqref{phih}. Using Eqs.~\eqref{phih} and \eqref{Gwf} we
receive for the vector $\mathbf{G}$ the following expression,
\begin{widetext}
\begin{equation}
  \label{GJphi}
  \mathbf{G}=\frac{\gamma}{2}
  \left\{
    \frac{1}{r^5}
    \left[
      r^2\mathbf{J}-3\mathbf{r}(\mathbf{J}\mathbf{r})
    \right]-
    \frac{\gamma}{\gamma+1}
    \left[
      \frac{1}{r^3}
      \left[
        v^2\mathbf{J}-\mathbf{v}(\mathbf{v}\mathbf{J})
      \right]-
      \frac{3}{r^5}
      [\mathbf{v}\times\mathbf{r}]
      (\mathbf{v}[\mathbf{r}\times\mathbf{J}])
    \right]+
    \frac{2\gamma+1}{\gamma+1}\frac{M}{r^3}
    [\mathbf{v}\times\mathbf{r}]
  \right\}.
\end{equation}
\end{widetext}
It can be seen that this expression for the vector $\mathbf{G}$
agrees with the analogous formula obtained in our prevoius work
\cite{DvoGriStu05} only for slow neutrinos having $v\ll 1$ (see
also above).  This discrepancy results from the incorrect account
of the spin-orbital interaction in Ref.~\cite{DvoGriStu05}.
However, if we study only the radial neutrino motion, the main
contribution to the neutrino spin evolution comes from the
interaction with the angular momentum of the massive object and we
reach an agreement with our recent work.

We consider therefore a neutrino propagating along the radius,
i.e. $\mathbf{v}=v(\mathbf{r}/r)$. In this case we can rewrite
Eq.~\eqref{GJphi} in the form,
\begin{equation}\label{GJradial}
  \mathbf{G}=\frac{\gamma}{2r^3}
  \left\{
    \frac{1}{\gamma}\mathbf{J}-
    \mathbf{n}(\mathbf{J}\mathbf{n})
    \left(
      2+\frac{1}{\gamma}
    \right)
  \right\},
\end{equation}
where $\mathbf{n}$ is the unit vector along the neutrino velocity.
One can see that the vector $\mathbf{G}$ given in
Eq.~\eqref{GJradial} coincides with that derived in our previous
work (see Ref.~\cite{DvoGriStu05}). Thus all results concerning
the neutrino spin light obtained in Ref.~\cite{DvoGriStu05} are
valid \emph{only} for the radial neutrino motion.

Now let us write down the effective Hamiltonian for neutrino spin
oscillations in our case. We can always choose the coordinate
system so that,
\begin{gather}
  \label{nJ}
  \mathbf{n}=(0,0,1),
  \quad
  \mathbf{J}=(J_1,0,J_3),
  \\
  \notag
  J_1=J\sin\vartheta,
  \quad
  J_3=J\cos\vartheta,
\end{gather}
where $\vartheta$ is the angle between vectors $\mathbf{n}$ and
$\mathbf{J}$. Using Eqs.~\eqref{Heffwf}, \eqref{GJradial} and
\eqref{nJ} we obtain the expression for the neutrino spin
oscillations effective Hamiltonian,
\begin{equation}\label{HeffJ}
  H_\mathrm{eff}=\frac{J}{r^3}
  \begin{pmatrix}
    \cos\vartheta & \sin\vartheta/(2\gamma) \\
    \sin\vartheta/(2\gamma) & -\cos\vartheta
  \end{pmatrix}.
\end{equation}
Here $\mathbf{r}=\mathbf{r}_0+\mathbf{v}t$, where $\mathbf{r}_0$
is the initial neutrino coordinate. It follows from
Eq.~\eqref{HeffJ} that transitions between different polarization
states of a neutrino are suppressed by the factor $1/\gamma$ which
is small for ultrarelativistic neutrinos. The most intensive
neutrino spin oscillations take place when a neutrino is
propagating in the direction perpendicular to the vector
$\mathbf{J}$ ($\vartheta=\pi/2$). There are no oscillations at all
when a neutrino is propagating along the vector $\mathbf{J}$
($\vartheta=0$).

The transition probability for the $\nu_L\leftrightarrow\nu_R$
oscillations can be obtained with help of the effective
Hamiltonian given in Eq.~\eqref{HeffJ}. Let us suppose that
$\nu_R(0)=0$ and $\nu_L(0)=1$, then we receive the expression for
the probability to find a right-handed neutrino in this neutrino
beam,
\begin{equation}\label{Probabr}
  P(r)=
  \frac{\sin^2\vartheta}{4\gamma^2\cos^2\vartheta+\sin^2\vartheta}
  \sin^2[\Phi(r)],
\end{equation}
where
\begin{align}\label{Phir}
  \Phi(r)= & \frac{J}{4\sqrt{\gamma^2-1}}
  \left(
    \frac{1}{r_0^2}-\frac{1}{r^2}
  \right)
  \\
  \notag
  & \times
  \sqrt{4\gamma^2\cos^2\vartheta+\sin^2\vartheta}.
\end{align}
Here $r_0$ is the distance between a neutrino and the center of
the massive object at $t=0$. Eqs.~\eqref{Probabr} and \eqref{Phir}
coincide with the similar expression for the transition
probability derived in our previous work \cite{DvoGriStu05}.

\section{Neutrino spin oscillations in Schwarzschild metric}
\label{SM}

General equations derived in Sec.~\ref{General} allow one to
describe particle's spin evolution not only in weak gravitational
fields. In this section we discuss the neutrino spin evolution and
oscillations when a particle interacts with a non-rotating black
hole in the case when a neutrino propagates even in the vicinity
of a black hole.

When we study the gravitational field of a non-rotating black
hole, the interval is known to be expressed with help of the
Schwarzschild metric,
\begin{equation}\label{schwarz}
  d\tau^2=A^{2}dt^2-A^{-2}dr^2-
  r^2(d\theta^2+\sin^2\theta d\phi^2),
\end{equation}
where
\begin{equation}\label{Aschwarz}
  A=\sqrt{1-\frac{r_g}{r}},
\end{equation}
and $r_g$ is the Schwarzschild radius. In Eq.~\eqref{schwarz} we
use the spherical coordinate system for the world (not for the
vierbein) coordinates. We can decompose the four-velocity $U^\mu$
with help of the spherical coordinates,
\begin{equation}\label{Umuspher}
  U^\mu=(U^0,U_r,U_\theta,U_\phi).
\end{equation}

In order to construct the "electric" and "magnetic" fields we
should choose the appropriate vierbein vectors. One can verify
that the following vectors,
\begin{align}
  \label{e0}
  e^{0}{}_{\mu} & =(A,0,0,0), \\
  e^{1}{}_{\mu} & =(0,A^{-1},0,0), \\
  e^{2}{}_{\mu} & =(0,0,r,0), \\
  \label{e3}
  e^{3}{}_{\mu} & =(0,0,0,r\sin\theta),
\end{align}
satisfy the general properties of the vierbein vectors given in
Eq.~\eqref{vierbeinprop}.

Using Eqs.~\eqref{Umuspher}-\eqref{e3} we can find vierbein
components of the vector $u^a$,
\begin{equation*}
  u^a=(\gamma A,U_r A^{-1},U_\theta r,U_\phi r\sin\theta).
\end{equation*}
With help of Eqs.~\eqref{cde0}-\eqref{Gabconv} as well as using
the definition of the tensor $G_{ab}$ one can write down the
components of the "electric" and "magnetic" fields,
\begin{align}\label{EBschwarz}
  \mathbf{E} & =
  \left(
    -\gamma\frac{r_g}{2r^2},0,0
  \right),
  \\
  \notag
  \mathbf{B} & =(U_\phi\cos\theta,-U_\phi A\sin\theta,U_\theta A).
\end{align}
Neutrino spin precession is determined by the vector
$\bm{\Omega}=\mathbf{G}/\gamma$. The components of this vector can
be found on the basis of Eqs.~\eqref{GEB} and \eqref{EBschwarz},
\begin{align}
  \label{Omega1}
  \Omega_1 & =\frac{1}{2}v_\phi\cos\theta, \\
  \label{Omega2}
  \Omega_2 & =v_\phi\sin\theta\frac{1}{2}
  \left(
    -A+\frac{\gamma}{(1+\gamma A)}
    \frac{r_g}{2r}
  \right), \\
  \label{Omega3}
  \Omega_3 & =v_\theta\frac{1}{2}
  \left(
    A-\frac{\gamma}{(1+\gamma A)}
    \frac{r_g}{2r}
  \right),
\end{align}
where $\mathbf{v}=(v_r,v_\theta,v_\phi)$ are the components of the
world (not the vierbein) velocity. In deriving of
Eqs.~\eqref{Omega1}-\eqref{Omega3} we use the fact that
$\mathbf{U}=\gamma\mathbf{v}$.

Let us discuss a neutrino orbiting a black hole. For simplicity we
consider only circular orbits with the radius $R$. We may restrict
ourselves to the studying of the orbits lying only in the
equatorial plane ($\theta=\pi/2$ or equivalently $v_\theta=0$)
because of the spherical symmetry of the gravitational field. In
this case $\Omega_1=0$ and $\Omega_3=0$ in Eqs.~\eqref{Omega1} and
\eqref{Omega3}. The expressions for the neutrino angular velocity
and $\gamma$ are presented in Ref.~\cite{Wei72p185},
\begin{equation}\label{vphico}
  v_\phi=\frac{d\phi}{dt}=\sqrt{\frac{r_g}{2R^3}},
\end{equation}
and
\begin{equation}\label{gammaco}
  \gamma^{-1}=\frac{d\tau}{dt}=\sqrt{1-\frac{3r_g}{2R}}.
\end{equation}
It should be noted that the vierbein four velocity now takes the
form, $u^a=(\gamma A, 0, 0, \gamma v_\phi r)$. One can verify that
$u^a u_a=1$ with help of Eqs.~\eqref{Aschwarz}, \eqref{vphico} and
\eqref{gammaco}. We also mention that Eq.~\eqref{uG} is also
identically satisfied. Indeed $G_{ab}u^b=0$ since $E_1 u^0=B_2
u^3$ and hence $du^a/d\tau=0$. Therefore a neutrino has constant
four velocity with respect to the vierbein frame. Using
Eqs.~\eqref{Aschwarz}, \eqref{Omega2} and \eqref{gammaco} we can
rewrite $\Omega_2$ in the more simple form,
\begin{equation}\label{Omaga2sf}
  \Omega_2=-\gamma^{-1}\frac{v_\phi}{2}.
\end{equation}
This equation has very clear physical meaning. It is well known
that the spin precession of a charged particle with
$\mathfrak{g}=2$ in the external electromagnetic field is
completely determined by the particle's motion law (see
Sec.~\ref{General} or Ref.~\cite{Zwa65}). Therefore
Eq.~\eqref{Omaga2sf} is nothing else as the Lorentz transformation
of the spin rotation frequency from the rest frame to the
laboratory frame.

We suppose that initially a neutrino is left-handed, i.e. its
initial spin vector is antiparallel to the particle's velocity.
According to Eqs.~\eqref{zetaeq} and \eqref{Omega1}-\eqref{Omega3}
the neutrino spin rotates around the second axis. Therefore using
Eq.~\eqref{zetaeq} we can construct the effective Hamiltonian for
the neutrino spin oscillations in the gravitational field of a
non-rotating black hole,
\begin{equation}\label{HeffBH}
  H_\mathrm{eff}=
  \begin{pmatrix}
    0 & -i\Omega_2 \\
    i\Omega_2 & 0
  \end{pmatrix}.
\end{equation}
It is interesting to note that in Eq.~\eqref{HeffBH} the parameter
$|\Omega_2|{\not=}0$ if $\gamma^{-1}{\not=}0$ (see below). One can
see it directly in Eq.~\eqref{Omaga2sf}. Using Eq.~\eqref{HeffBH}
we write the expression for the neutrino transition probability,
\begin{equation}\label{probBH}
  P(t)=\sin^2(\Omega_2 t).
\end{equation}
We can see that there is a full mixing in our case and neutrino
transition probability can achieve a unit value.

It is possible to describe the neutrino spin evolution using
formalism elaborated in Ref.~\cite{LobPav99} where the the
approach for the studying of Lorentz and Bargmann-Michel-Telegdi
equations was described. It follows from the results of that work
that the spin dynamics is completely determined by the particle's
motion law when $\mathfrak{g}=2$. Eq.~\eqref{uG} can be rewritten
in the equivalent form,
\begin{equation*}
  \frac{d}{d\tau}\undertilde{\Lambda}=
  \undertilde{F}\undertilde{\Lambda},
  \quad
  \undertilde{F}=\frac{i}{2}\bm{\sigma}(\mathbf{B}-i\mathbf{E}),
\end{equation*}
and the operator $\undertilde{\Lambda}$ implements the shift of
the particle's velocity along the trajectory. We present the main
properties of the $SL(2,C)$ group in Appendix~\ref{LorGroup}. If
one has found the operator $\undertilde{\Lambda}$, the resolvent
of the Eq.~\eqref{sG} takes the form ($\mathfrak{g}=2$),
\begin{equation}\label{utRdef}
  \undertilde{R}(\tau,\tau_0)=
  \undertilde{L}^{-1}(\tau)\undertilde{\Lambda}\undertilde{L}(\tau_0),
\end{equation}
where the matrix $\undertilde{L}$ implements the boost from the
rest frame to the laboratory frame.

If we consider the neutrino motion in the equatorial plane from
the very beginning, the operator $\undertilde{\Lambda}$ is
expressed in the following way,
\begin{equation}\label{utLambda}
  \undertilde{\Lambda}=\exp(\undertilde{F}\tau)=
  \begin{pmatrix}
    \cos\alpha t & \alpha_{+}\sin\alpha t \\
    \alpha_{-}\sin\alpha t & \cos\alpha t
  \end{pmatrix},
\end{equation}
where $\alpha=\sqrt{B^2-E^2}/(2\gamma)$ and
\begin{equation*}
  \alpha_{\pm}=\frac{E\pm B}{\sqrt{B^2-E^2}}.
\end{equation*}
Note that using Eqs.~\eqref{Aschwarz}, \eqref{EBschwarz},
\eqref{vphico} and \eqref{gammaco} we can show that
$\alpha=v_\phi/(2\gamma)$.
It can be verified that
$\undertilde{\Lambda}\undertilde{u}\undertilde{\Lambda}^\dagger=
\undertilde{u}$, where
\begin{equation*}
  \undertilde{u}=
  \begin{pmatrix}
    u^0+u^3 & 0 \\
    0 & u^0-u^3
  \end{pmatrix},
\end{equation*}
i.e. a neutrino moves along a strait line (in the veirbein frame).
This fact also agrees with the result obtained above in present
paper. The operator $\undertilde{L}$ does not depend on time since
the neutrino's four velocity is constant. This matrix is expressed
in the following way (see Appendix~\ref{LorGroup}),
\begin{equation}\label{utL}
  \undertilde{L}=
  \begin{pmatrix}
    \sqrt{u^0+u^3} & 0 \\
    0 & \sqrt{u^0-u^3}
  \end{pmatrix}.
\end{equation}
Using Eqs.~\eqref{utRdef}-\eqref{utL} we receive the resolvent
$\undertilde{R}(t)$ in the form (here we assume that $t_0=0$),
\begin{equation}\label{utR}
  \undertilde{R}(t)=
  \begin{pmatrix}
    \cos\alpha t & \sin\alpha t \\
    -\sin\alpha t & \cos\alpha t
  \end{pmatrix}.
\end{equation}
In order to find the spin rotation axis one adopts the unit vector
$\mathbf{k}$ along this axis, so that
$[\undertilde{R},\bm{\sigma}\mathbf{k}]=0$. Here use the fact that
$\undertilde{R}\in SO(3)$ and
$\undertilde{R}^{\dagger}=\undertilde{R}^{-1}$. It can be easily
derived with help of Eq.~\eqref{utR} that
$[\undertilde{R},\sigma_y]=0$. Thus the spin precession axis is
the second axis in the complete agreement with the above obtained
results.

\begin{figure}
  \centering
  \includegraphics[scale=.45]{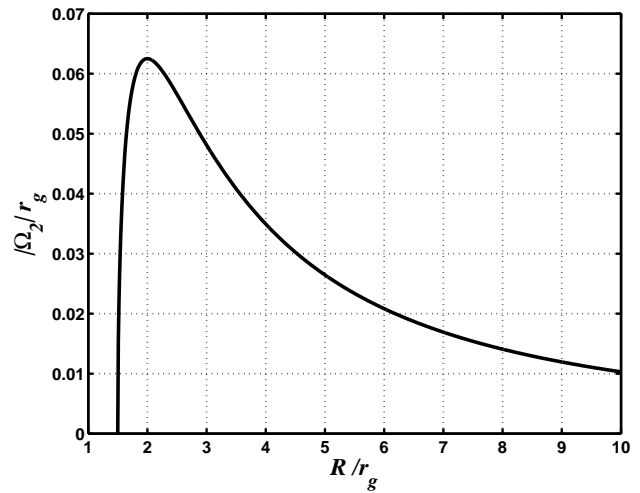}
  \caption{Neutrino spin oscillations frequency versus the radius
  of the neutrino orbit.}\label{freq}
\end{figure}

One can also plot the frequency of neutrino spin oscillations
versus the radius of the orbit. It is possible to see on
Fig.~\ref{freq} that $|\Omega_2|=0$ at $R=1.5 r_g$ and
$|\Omega_2|\to 0$ at $R\to\infty$.
One can conclude that $|\Omega_2|$ has its maximal value (equal to
$6.25\times 10^{-2} r_g^{-1}$) at $R=2 r_g$ [see also
Eq.~\eqref{Omaga2sf}]. Let us evaluate the number of revolutions
$N$, that a neutrino should make, necessary for the total spin
flip. Using Eqs.~\eqref{vphico} and \eqref{Omaga2sf} we can easily
find for $N$,
\begin{equation*}
  N=\frac{v_\phi}{2|\Omega_2|}=
  \left.
    \gamma
  \right|_{R=2 r_g}
  =2.
\end{equation*}
It is interesting to evaluate the characteristic period of the
neutrino spin oscillations (oscillations length). For $M=10
M_\odot$ at $R=2 r_g$ we get for $T=\pi/|\Omega_2|\approx
4.94\times 10^{-3}\thinspace\text{s}$.

We can discuss another orbit, namely $R=1.5 r_g$. It follows from
Eq.~\eqref{Omaga2sf} (see also Fig.~\ref{freq}) that the frequency
of the transitions between two polarization states vanishes. It
should be noted that $d\tau=0$ at $R=1.5 r_g$ [see
Eq.~\eqref{gammaco}]. Therefore this orbit corresponds to massless
particles. We reveal that the property of a massless particle in
the Minkowsky space-time to keep unchanged its spin direction with
respect to the momentum remains unaffected in the Schwarzschild
metric. It is interesting to note that the absence of the spin
rotation for massless particles is valid only for the
Schwarzschild metric. It was shown in
Refs.~\cite{MohMukPra02,DebMukDad06} that for a background metric
with at least one off-diagonal space component, e.g. the Kerr
metric, in the limit $m_\nu \to 0$ still there are transitions
between neutrino and antineutrino, i.e. neutrino spin flip. This
phenomenon is used in Refs.~\cite{MohMukPra02,Muk05} to explain
the neutrino-antineutrino asymmetry problem.

It is also worth examining the behaviour of $\Omega_2$ for large
orbits. If a neutrino moves on a remote orbit, the gravitational
field is weak. Thus we again do not distinguish between world and
vierbein indexes. We define the coordinate system so that the
angular velocity is expressed in the following way,
$\bm{\omega}_\mathrm{frame}=v_{\phi}\mathbf{e}_z$, where
$\mathbf{e}_z$ is the unit vector. Using Eq.~\eqref{Omaga2sf} in
the weak field limit we obtain for the spin rotation frequency,
\begin{equation*}
  \bm{\omega}_\mathrm{abs}=-2\Omega_2\mathbf{e}_z=v_\phi
  \left(
    1-\frac{3}{4}\frac{r_g}{R}+\dots
  \right)\mathbf{e}_z,
\end{equation*}
since the second vierbein coordinate axis is nothing else as the
$\theta$-axis and for the orbit in the equatorial plane we have
$\mathbf{e}_\theta=-\mathbf{e}_z$. Note that the effective
Hamiltonian in Eq.~\eqref{HeffBH} determines the evolution of the
neutrino rest-frame-spin vector. However an observer is in the
laboratory frame ($t$ is time in the laboratory frame). Thus
$\Omega_2$ is the oscillations frequency measured in the
laboratory frame. In order to obtain the spin rotation frequency
measured by an observer in the co-moving frame we should subtract
$\bm{\omega}_\mathrm{frame}$ from $\bm{\omega}_\mathrm{abs}$ since
$\bm{\omega}_\mathrm{frame}$ is the angular velocity of the
orbital motion, i.e. the angular velocity of the rest frame
rotation with respect to the laboratory frame. Finally we get for
$\bm{\omega}_\mathrm{rel}=\bm{\omega}_\mathrm{abs}-
\bm{\omega}_\mathrm{frame}$ the following expression,
\begin{equation*}
  \bm{\omega}_\mathrm{rel}=
  -\frac{3}{4}\frac{r_g}{R} v_\phi \mathbf{e}_z,
\end{equation*}
which agrees with the classical result for the spin rotation
frequency in the co-moving frame (see, e.g., Ref.~\cite{Sch60} or
Eq.~\eqref{GJphi} in the limit $\gamma\to 1$).

At the end of this section we briefly examine the applicability of
the quasi-classical approach to the description of the neutrino
spin evolution in the gravitational field of a black hole. Basing
on the similarity between particle spin precession in
gravitational and electromagnetic fields we can apply the results
of Ref.~\cite{Ter90} where the quasi-classical approximation for
the electron spin precession in an external magnetic field was
considered. It was found in that paper that one can use the
quasi-classical limit if the corresponding inequality (we rewrite
it for our purposes) is satisfied,
\begin{equation}\label{qqlcond}
  \frac{\hbar}{2\mathcal{E}}
  \left|
    \frac{d\bm{\zeta}}{dt'}
  \right|
  \ll 1
\end{equation}
where $\mathcal{E}$ is the neutrino energy in the vierbein frame,
the derivative is also taken with respect to the vierbein time
$t'$. We again study the case of circular orbits. For the neutrino
energy one finds, $\mathcal{E}=m_\nu u^0=m_\nu \gamma A$, where
$m_\nu$ is the neutrino mass. Using Eq.~\eqref{e0} we obtain that
$dt'=A dt$. Let us discuss an electron neutrino with mass
$m_{\nu_e}\approx 2\thinspace\text{eV}$ rotating around a black
hole with $M=10 M_\odot$ on the orbit with radius $R=2r_g$. Then
Eq.~\eqref{qqlcond} reads,
\begin{equation*}
  \frac{\hbar}{2\mathcal{E}}
  \left|
    \frac{d\bm{\zeta}}{dt'}
  \right|=\frac{\hbar}{2m_\nu \gamma^2 A^2}v_\phi\approx
  4.19\times 10^{-13}\ll 1.
\end{equation*}
We can see that the condition of the validity of the
quasi-classical approach is satisfied almost in all reasonable
astrophysical objects.

\section{Conclusion}\label{CONCL}

In conclusion we note that neutrino spin oscillations in
gravitational fields within the quasi-classical approach have been
studied. We have started from the particle spin evolution equation
in the vierbein frame. On the the basis of this approach we have
investigated neutrino spin oscillations in a weak gravitational
field created by a massive rotating object. The three dimensional
spin evolution equation has been obtained. We have improved
analogous calculation performed in our previous work
\cite{DvoGriStu05}. The neutrino spin evolution equation derived
in the present work is valid for arbitrary neutrino velocities and
correctly takes into account the neutrino velocity dependence.
Then we have written down the neutrino oscillations effective
Hamiltonian [Eq.~\eqref{HeffJ}] and transition probability
[Eqs.~\eqref{Probabr} and \eqref{Phir}] when a neutrino propagates
along the radial direction. The neutrino spin conversion rate has
been analyzed for different neutrino velocity directions. Then we
have examined neutrino spin oscillations in the gravitational
field of a non-rotating black hole. We have studied a neutrino
moving on circular orbits, with our calculations being valid for
the neutrino motion in the vicinity of a black hole. We have
derived the effective Hamiltonian [Eq.~\eqref{HeffBH}] and
transition probability [Eq.~\eqref{probBH}] in this case. It has
been demonstrated that neutrino spin oscillations occur in the
Schwarzschield metric. We have also analyzed the dependence of the
neutrino spin oscillations frequency on the radius of the orbit.
The validity of the quasi-classical approach to the description of
neutrino spin oscillations in a gravitational field has been
analyzed in our work.

It should be noted that the method used in the present paper for
the studying of neutrino spin oscillations is valid not only for
neutrinos but also for any spin-$1/2$ particles. The basic
equations \eqref{zetaeq} and \eqref{GEB} were demonstrated in
Ref.~\cite{SilTer05} to be applicable to a massive Dirac particle
in a weak static gravitational field. Therefore we can study,
e.g., electron spin evolution with help of this approach. It is
known that rather frequently massive gravitational objects have
inherent magnetic fields and are surrounded by the moving matter
(accretion disks). Thus, if we discuss the particle spin
oscillations process comprehensively, i.e. take into account all
factors, the cases of neutrinos and electrons are different
because these particles have diverse types of the interaction with
magnetic field and with the background matter. In principle
massless particles can be also included into consideration (see
Sec.~\ref{SM}). Despite of the fact that the mass of a particle is
absent in Eqs.~\eqref{zetaeq} and \eqref{GEB}, the law of motion
(i.e. the function $u^a(\mathbf{r},t)$, which is involved in the
spin evolution equations) of massless particles differs from that
of massive particles. However in the case of massless particles
the helicity and chirality are the same and one can apply the
methods used in the recent work~\cite{Muk05}.

According to Eqs.~\eqref{Omega1}-\eqref{Omega3} there is no
neutrino spin flip in case of radial neutrino motion in the
Schwarzschield metric. This our result agrees with the deduction
of Ref.~\cite{PirRoyWud96} where the radial propagation of
neutrinos from active galactic nuclei was studied. However one can
hardly agree with the claims of Ref.~\cite{CarFul97} that a
spherically symmetric, static Schwarzschild space-time cannot
cause the particle spin flip. That statement confront both our
results and the analysis of Ref.~\cite{SilTer05} in which it was
shown by means of direct calculations that in the Schwarzschield
metric the spin of a Dirac fermion can precess and change its
direction with respect to the particle momentum. The calculations
in that paper were based on Dirac equation in curved space. That
result also agrees with the statement of our paper that the
gravitational field of a non-rotating black hole can cause
neutrino spin oscillations.

Note that the results of our paper can be applied to the
description of the neutrino spin evolution in various
astrophysical media. It is known that relic massive neutrinos can
cluster into a halo around a galaxy contributing to cold dark
matter. Despite of the fact that the average size of a halo ranges
between $10^{2}$ and $10^{3}\thinspace\text{kpc}$ for a typical
galaxy like the Milky Way (see, e.g., Ref.~\cite{RinWon04}), there
is a possibility for a neutrino to be gravitationally captured on
orbits close to the central massive object. In this case strong
gravitational field can essentially contribute to the neutrino
spin oscillations process.

\begin{acknowledgments}
This research was supported by grant of Russian Science Support
Foundation. The author is very thankful to Andrey Lobanov (MSU)
and Andrey Zorin (JINR) for helpful discussions as well as to
Takuya Morozumi for making the facilities of Hiroshima University
available for him and for hospitality. The referee's comments are
also appreciated.
\end{acknowledgments}

\appendix

\section{Covariant derivatives of the vierbein vectors in
Schwarzschild metric}\label{VF}

The covariant derivatives of the veirbein vectors enter in the
definition of the tensor $G_{ab}$. In this appendix we present the
main formulae necessary for the covariant derivatives calculation.
The covariant derivative of $e_{a \mu}$ is equal to
\begin{equation}\label{covderdef}
  e_{a\mu;\nu}=
  \frac{\partial e_{a\mu}}{\partial x^\nu}-
  \Gamma^{\lambda}{}_{\mu\nu}e_{a\lambda},
\end{equation}
where $\Gamma^{\lambda}{}_{\mu\nu}$ are the Christoffel symbols.
The non-zero Christoffel symbols for the Schwarzschield metric are
(see, e.g., Ref.~\cite{LanLif88p387}),
\begin{widetext}
\begin{gather}\label{ChristSch}
  \Gamma^{r}{}_{rr}= \frac{r_g}{2r(r_g-r)},
  \quad
  \Gamma^{r}{}_{\theta\theta}= r_g-r,
  \quad
  \Gamma^{r}{}_{tt}= \frac{r_g(r_g-r)}{2r^3},
  \quad
  \Gamma^{\theta}{}_{r\theta}= \frac{1}{r},
  \quad
  \Gamma^{\phi}{}_{\phi r}= \frac{1}{r},
  \\
  \Gamma^{\phi}{}_{\phi\theta}= \cot\theta,
  \quad
  \Gamma^{\theta}{}_{\phi\phi}= -\sin\theta\cos\theta,
  \quad
  \Gamma^{r}{}_{\phi\phi}= (r_g-r)\sin^2\theta,
  \quad
  \Gamma^{t}{}_{tr}= \frac{r_g}{2r(r_g-r)}.
  \notag
\end{gather}
\end{widetext}
Using Eqs.~\eqref{e0}-\eqref{e3}, \eqref{covderdef} and
\eqref{ChristSch} we obtain the expressions for the covariant
derivatives in the following form,
\begin{equation}\label{cde0}
  e_{0\mu;\nu}=
  \begin{pmatrix}
    0 & 0 & 0 & 0 \\
    -(r_g/2r^2)A^{-1} & 0 & 0 & 0 \\
    0 & 0 & 0 & 0 \\
    0 & 0 & 0 & 0
  \end{pmatrix},
\end{equation}
\begin{equation}
  e_{1\mu;\nu}=
  \begin{pmatrix}
    (r_g/2r^2)A & 0 & 0 & 0 \\
    0 & 0 & 0 & 0 \\
    0 & 0 & (r_g-r)A^{-1} & 0 \\
    0 & 0 & 0 & -r\sin^2\theta
  \end{pmatrix},
\end{equation}
\begin{equation}
  e_{2\mu;\nu}=
  \begin{pmatrix}
    0 & 0 & 0 & 0 \\
    0 & 0 & 1 & 0 \\
    0 & 0 & 0 & 0 \\
    0 & 0 & 0 & -r\sin\theta\cos\theta
  \end{pmatrix},
\end{equation}
\begin{equation}\label{cde3}
  e_{3\mu;\nu}=
  \begin{pmatrix}
    0 & 0 & 0 & 0 \\
    0 & 0 & 0 & \sin\theta \\
    0 & 0 & 0 & r\cos\theta \\
    0 & 0 & 0 & 0
  \end{pmatrix},
\end{equation}
where the first index ($\mu$) numbers lines and the second one
($\nu$) numbers the rows of the matrix.

It is convenient to rewrite the definition of the tensor $G_{ab}$
in the following way,
\begin{equation}\label{Gabconv}
  G_{ab}=e_{a\mu;\nu}e_{b}{}^{\mu}U^\nu.
\end{equation}
Here we used the properties of the vierbein vectors [see
Eq.~\eqref{vierbeinprop}]. Eq.~\eqref{Gabconv} is more convenient
for the further calculations since it allows one to express the
tensor $G_{ab}$ directly in terms of the components of the four
vector $U^\nu$ rather than $u^a$.

\section{Universal covering of the Lorentz group}\label{LorGroup}

It is well known that $SL(2,C)$ is the universal covering group
for the orthochronous proper Lorentz subgroup $L^{\uparrow}_{+}$
($\det\Lambda=1$ and $\Lambda^{0}{}_{0}>1$ for $\Lambda\in
L^{\uparrow}_{+}$). In this appendix we briefly discuss the local
isomorphism between these groups (see also
Ref.~\cite{BogLogOksTod87}).

For any element $x^\mu$ of the minkowskian space we set the matrix
$\undertilde{x}=x^\mu\undertilde{e}_\mu$, where $\undertilde{e}_0$
is the $2\times 2$ unit matrix and
$\undertilde{\mathbf{e}}=\bm{\sigma}$. The Lorentz transformation
acts on $\undertilde{x}$ as $\undertilde{\Lambda x}=
\undertilde{\Lambda}\undertilde{x}\undertilde{\Lambda}^\dagger$,
where $\undertilde{\Lambda}^\dagger$ is the Hermitian cojugate to
the matrix $\undertilde{\Lambda}$. For instance, the Lorentz
transformation of the form,
\begin{align*}
  y^0 = & x^0\cosh\chi+
  (\mathbf{x}\mathbf{l})\sinh\chi,
  \\
  \mathbf{y} = & \mathbf{x}-
  (\mathbf{x}\mathbf{l})\mathbf{l}+
  [(\mathbf{x}\mathbf{l})\cosh\chi+
  x^0\sinh\chi]\mathbf{l},
\end{align*}
corresponds to the matrix
$H(\mathbf{l},\chi)=\exp[\chi(\bm{\sigma}\mathbf{l})/2]$. Here
$\mathbf{l}$ is the three dimensional unit vector and $\chi$ is
the real parameter.

\bibliography{generaleng}

\end{document}